\documentclass[11pt]{article}
\usepackage{geometry}                
\usepackage{graphicx}
\usepackage{xcolor}
\usepackage{bbm}
\usepackage{cite}
\newcommand{\be}{\begin{equation}}
\newcommand{\ee}{\end{equation}}
\newcommand{\bea}{\begin{eqnarray}}
\newcommand{\eea}{\end{eqnarray}}
\newcommand{\beas}{\begin{eqnarray*}}
\newcommand{\eeas}{\end{eqnarray*}}
\newcommand{\ER}{Erd\H{o}s-R\'enyi }
\usepackage{amssymb}
\usepackage{amsmath}
\usepackage{amsfonts}
\usepackage{graphicx}
\usepackage{epstopdf}
\usepackage{array} 
\usepackage{booktabs} 
\usepackage{multirow}
\usepackage[utf8]{inputenc}

\title{Understanding Financial Contagion: A Complexity Modeling Perspective\thanks{This article will be a contributed chapter to the SFI edited volume: The Economy as a Complex Evolving System, Part IV}}

\author{Fabio Caccioli\thanks{Department of Computer Science, University College London,  London WC1E 6EA, United Kingdom. Systemic Risk Centre, London School of Economics and Political Science, London WC2A 2AE, United Kingdom.}}

\begin{document}
\maketitle

\abstract{This chapter reviews key contributions of complexity science to the study of systemic risk in financial systems. The focus is on network models of financial contagion, where I explore various mechanisms of shock propagation, such as counterparty default risk and overlapping portfolios. I highlight how the interconnectedness of financial institutions can amplify risk, and I discuss how standard risk management tools, which neglect these interactions, can increase systemic risk.}

\section{Introduction}

The Global Financial Crisis of $2008$ made it clear that, in addition to managing traditional risks such as { market risk}---the risk of adverse fluctuations in the value of investments due to market movements---{credit risk}---the risk of potential default by counterparties---and {liquidity risk}---the risk of being unable to meet short-term financial obligations--- a new kind of risk needs to be managed: {systemic risk}. 

Systemic risk refers to the risk associated with the collapse of the financial system as a whole, rather than the failure of individual institutions. Unlike other categories of risk, systemic risk is a collective property of the system, emerging from the interactions and interconnections between the many heterogeneous players operating in financial markets. Even more concerning, it can arise as an unintended consequence of agents who are attempting to reduce their own individual risks using traditional risk management tools. 

The study of systemic risk is of obvious importance to regulators, whose role is to ensure the stability of the financial system and the broader economy. In fact, during his opening address at the ECB Central Banking Conference in Frankfurt on 18 November 2010, the then-president of the ECB Jean-Claude Trichet famously stated  \cite{trichet2010}, ``I found the available models of limited help. In fact, I would go further: in the face of the crisis, we felt abandoned by conventional tools.''. However, understanding how systemic risk builds up in the system is equally important for individual institutions. As David Viniar, Goldman Sachs' Chief Financial Officer, reported in 2007: ``We were seeing things that were 25-standard deviation moves, several days in a row.'' \cite{financialtimes2007}. Quotes like these, though anecdotal, make it clear that standard risk management tools, which may work well in normal market conditions, fail under certain market regimes, and that new tools to complement traditional approaches are required.

These new tools need to explicitly account for the interactions and interconnections within the financial system, modeling the buildup of systemic risk from the bottom up. Unlike traditional tools that often view institutions in isolation, effective models of systemic risk must consider how risks propagate through the network of players and how seemingly isolated events can cascade into system-wide crises. This approach requires a fundamental shift in risk management, one that embraces complexity and interdependence as core features of the financial system.

Given that systemic risk is a collective property that emerges from the interactions between agents, it is particularly well-suited to be studied through the lens of complex systems science \cite{haldane2011systemic}.

In this chapter, I will review some of the work on modeling systemic risk using tools from complex systems science. This review will not be exhaustive, as the field of systemic risk is highly interdisciplinary and has been addressed using a variety of approaches. Instead, I will highlight key contributions that, in my view, complexity science has brought to the field.

\section{Financial networks}

For more than $20$ years, {networks} have been one of the primary tools used in the modeling of complex systems. A network is simply a mathematical object composed of {nodes}, which represent the components of a system, and {links}, which connect some of these nodes. As such, networks provide a natural way to encode the structure of interactions between the units that make up a complex system.

The contributions of network science to the study of complex systems are vast, spanning fields such as biological systems \cite{jeong2000large}, ecological systems \cite{dunne2006network}, transportation networks \cite{guimera2005worldwide}, technological networks \cite{faloutsos1999power}, and, of course, social networks \cite{liljeros2001web}, to name a few.

As mentioned earlier, the financial system is also composed of many interacting units. Some of these interactions, such as lending relationships between financial institutions, can naturally be modeled using networks.

In financial networks, nodes represent financial institutions (such as banks, funds, insurance companies), and links represent the relationships between them. Financial institutions interact in a variety of ways: lending money to each other, trading derivatives, forming ownership ties, or investing in similar products. Some relationships are direct, such as when one bank lends money to another under a formal contract. Others are indirect, like when two institutions are connected through the market because they invest in the same financial products.

These relationships evolve over time as old contracts come to maturity, new ones are established, and portfolios are rebalanced.
A comprehensive description of the financial system would therefore be in terms of a multilayer dynamical network, where each layer represents a specific type of relationships, and where each layer changes over time.

While dynamical multilayer modeling has become feasible in recent years \cite{bargigli2015multiplex,poledna2015multi}, historically, most research has focused on individual static layers due to limited data and the need to begin with simpler models.

Due to regulatory boundaries, much of the empirical research on financial networks has focused on national interbank systems (see e.g. \cite{boss2004network,soramaki2007topology,de2006fitness,iazzetta2009topology,cont2013network,craig2014interbank,fricke2015core}). Collectively, these studies have revealed several general features common across different networks, much like stylized facts in financial markets. Specifically, interbank networks exhibit the following characteristics:

\begin{itemize}
\item {\bf Heavy-tailed degree distributions:} A node's degree is the number of its links, which in interbank networks corresponds to the number of counterparties. Unsurprisingly, these systems are characterized by a few highly connected hubs (which can often be identified as systemically important institutions) and many nodes with low connectivity.
\item {\bf High clustering:} In network science, clustering refers to how interconnected a node's neighbors are, often measured by the clustering coefficient. This measures the abundance of triangles (closed loops) around a node. Like many real-world networks, interbank networks have a high level of clustering compared to random networks, meaning that banks tend to have highly interconnected counterparts.
\item {\bf Negative assortativity:}  Assortativity measures the tendency of nodes to be connected with others of similar degree. In social networks, positive assortativity is common, where high-degree nodes are connected to other high-degree nodes. However, in interbank networks, negative assortativity is observed---highly connected nodes tend to be linked with less-connected ones, similar to technological networks such as the internet.
\item {\bf Core-periphery structure:} These networks often feature a core of highly interconnected nodes (typically the systemically important institutions), surrounded by a periphery of less connected institutions.
\end{itemize}

When studying financial networks, a key goal is to develop models that describe how shocks propagate between institutions. If a bank or group of banks fails, its neighbors may be affected and potentially fail as well, setting off a chain reaction. The central question is: under what conditions does an initial shock escalate into a full-blown crisis, leading to the collapse of the network? Understanding the structural properties of interbank networks is critical, as the system's topology directly influences how shocks propagate.

Financial contagion refers to the propagation of shocks between financial institutions. The term is borrowed from epidemiology, and the analogy is intuitive: just as an infected individual can spread a disease to others they interact with, a stressed financial institution can transmit shocks to those it is connected with.

As mentioned earlier, financial institutions interact in various ways, and these interactions give rise to different contagion mechanisms and channels. These mechanisms can be classified along different dimensions, but a key distinction is between direct vs. indirect contagion and solvency vs. liquidity contagion.
\begin{itemize}
\item {\bf Direct contagion} occurs when there is an explicit relationship, such as an interbank loan, between two institutions.
\item {\bf Indirect contagion} arises when two institutions are connected through shared investments in the same assets. For example, if one institution is under stress and liquidates its position, the resulting drop in asset prices can cause losses for the other institution.
\end{itemize}

Similarly, contagion can also be characterized by the type of shock being transmitted:

\begin{itemize}
\item {\bf Solvency contagion} occurs when shocks propagate from the borrower to the lender. In the case of interbank loans, if the borrower defaults, the lender suffers losses because it is unable to recover the funds lent out. This is often referred to as counterparty default risk.
\item {\bf Liquidity contagion} occurs when shocks flow in the opposite direction, from lenders to borrowers. A typical example is a bank run, where depositors withdraw their money, causing the bank to default due to insufficient liquidity to meet all withdrawal demands.
\end{itemize}

In the following, I will discuss some results obtained for direct and indirect solvency contagion.

\section{Contagion due to counterparty default risk}\label{sec:counterpartyRiskContagion}

Contagion due to counterparty default risk occurs when an institution incurs a loss because a counterparty it is exposed to defaults or faces an increased risk of default. For simplicity, I will focus on interbank loans, where the lender is exposed to the borrower and suffers a loss if the borrower defaults. I will also refer primarily to banks throughout the chapter, though most models can be extended to other types of financial institutions. 

\begin{figure}[h]     \centering
    \includegraphics[width=1\textwidth]{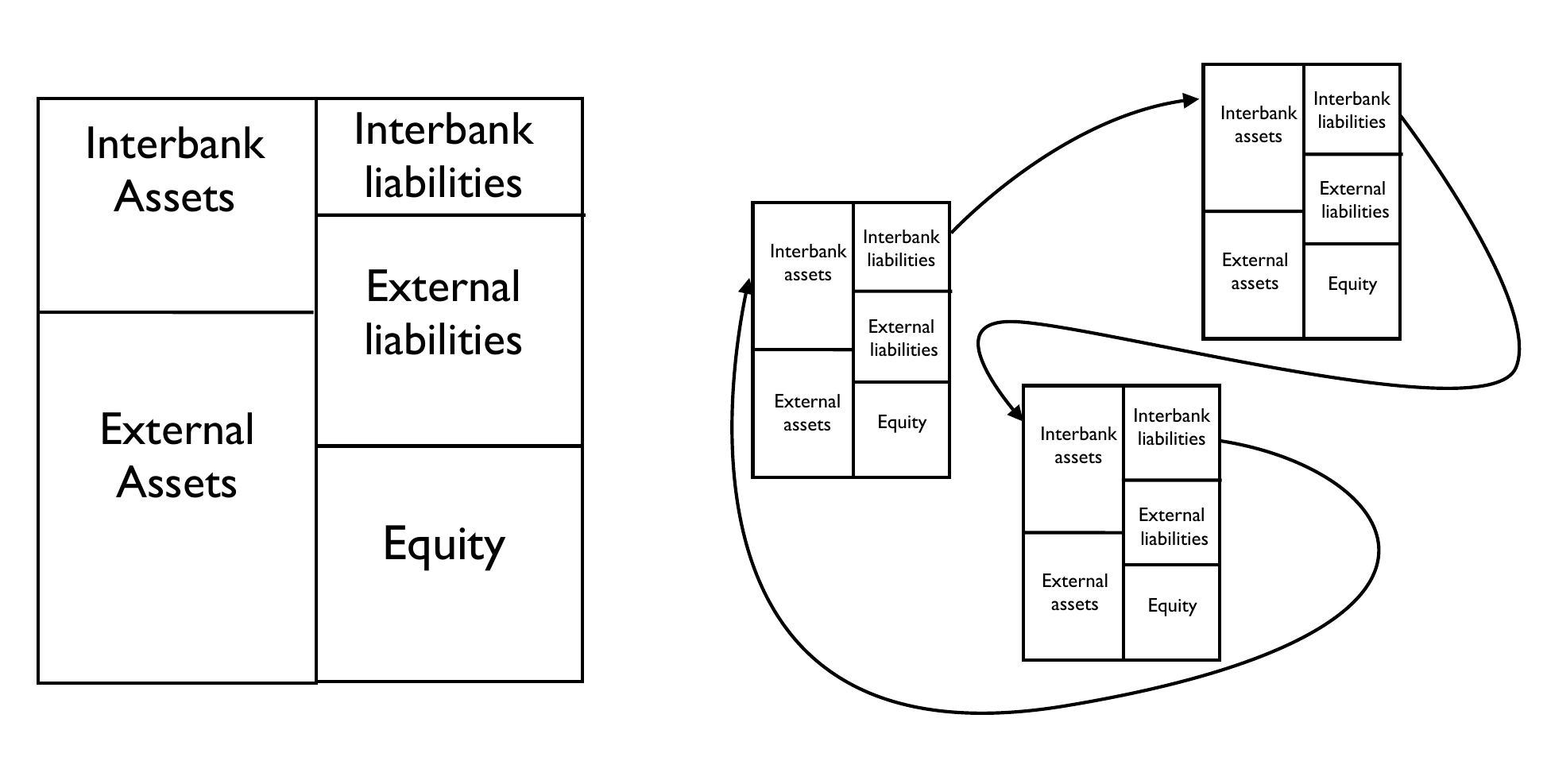}
        \caption{\footnotesize {\bf Left panel:} Stylized representation of balance sheet. Assets and liabilities are split into interbank and external. The value of assets equals that of liability plus equity. {\bf Right panel:} pictorial representation of a network of interconnected balance sheets. Shocks can propagate from borrowers to lenders through the links.}
    \label{fig:stylizedBalanceSheet} 
\end{figure}

A stylized representation of a bank can be described by its balance sheet (see left panel of figure \ref{fig:stylizedBalanceSheet}), which consists of an asset side (containing investments with positive value) and a liability side (containing debt). On the asset side, we distinguish between external assets and interbank assets. Interbank assets represent exposures to other banks in the network, such as interbank lending, while external assets are investments outside the network, such as stocks, bonds, and other securities.

Similarly, on the liability side, we differentiate between interbank liabilities and external liabilities. A bank's interbank assets are the interbank liabilities of other banks, meaning the network consists of interconnected balance sheets (see the right panel of figure \ref{fig:stylizedBalanceSheet} for a pictorial representation). The liability side also includes equity, which is the difference between assets and liabilities and represents the value owned by the bank's shareholders. Equity is the portion that remains after a bank liquidates all its assets and pays off its debt.

Mathematically, this relationship is expressed by the balance sheet identity:
\be
A_i^{\rm int}+A_i^{\rm ext} = L_i^{\rm int}+L_i^{\rm ext} +E_i,
\ee
where \(A_i^{\rm int}\) and \(A_i^{\rm ext}\) are the interbank and external assets of bank \(i\),  \(L_i^{\rm int}\) and \(L_i^{\rm      ext}\) are its interbank and external liabilities, and  \(E_i\) is the equity of bank \(i\).

A shock to a bank corresponds to a devaluation of its assets, and, according to the balance sheet identity, this must result in a reduction on the liability side. Since debt obligations must be honored, the equity absorbs losses first. If a bank's losses exceed its equity, even selling its entire assets would not cover its debt, rendering the bank insolvent. While technical default occurs when a debtor misses a payment, insolvency is often used as a proxy for default in financial contagion modeling.

Now, let's assume that when a bank defaults, its creditors lose the interbank assets associated with that bank. This is how contagion occurs: the default of one bank causes losses for its creditors, who may then default if the losses are large enough, potentially leading to a cascading effect throughout the network. This threshold contagion model is commonly referred to as Furfine's algorithm \cite{furfine2003interbank}, and, with some variations, it has been widely adopted in the literature.

I focus here on the work of Gai and Kapadia \cite{gai2010contagion}, who theoretically studied this dynamic on \ER random networks and derived conditions under which the initial default of a randomly selected bank could lead to a global cascade of defaults.

I highlight this paper because the model's behavior can be understood in terms of a percolation problem, making it an excellent example of how complex systems contribute to the understanding of systemic risk.

Gai and Kapadia consider a system of $N$ banks with the same amount of interbank assets, total assets, and equity. They model the system using directed \ER networks, where each potential interbank loan between any two directed pairs of banks is present with a probability $p$.

The probability $p$ is related to the average degree $\langle k \rangle$ of the network. The degree $k_i$ of node $i$ is the number of its links, and the average degree is the mean number of connections per node across the network $\langle k \rangle=\sum_i k_i/N$. Since the network is directed, we can further distinguish between in-degree ($k_i^{\rm in}$) and out-degree ($k_i^{\rm in}$), that are the number of incoming or outgoing links of node $i$ respectively. 

Gai and Kapadia assume in their model that the interbank assets of a bank $i$ are uniformly spread across its counterparties, so that each borrows an amount $A_i^{\rm int}/k_i^{\rm in}$. Here a link from $j$ to $i$ means that $j$ borrowed from $i$, i.e. the direction of the links indicates the propagation of losses. 

If we denote by $\sigma_i$ the state of bank i --- with $\sigma_i=1$ meaning default, and  $\sigma_i=0$ meaning that the bank is running --- the loss $\ell_i$ experienced by bank $i$ because of the failure of its counterparties is equal to

\be \label{eq:loss}
\ell_i = \sum_j \frac{A_{i}^{\rm int}}{k_i^{\rm in}}\sigma_j,
\ee
from which we can write the condition for the default as $i$ as

\be 
\sigma_i =
\begin{cases} 
1 & \text{if } \frac{1}{k_i^{\rm in}}\sum_j  \sigma_j >\frac{E_i}{A_{i}} \\
0 & \text{otherwise}
\end{cases}.
\ee

This dynamic mirrors Watts's model of global cascades in random networks \cite{watts2002Simple}, here generalized to directed networks. In fact, many financial contagion dynamics can be seen as variations of linear threshold models, where a node's state depends on the states of its neighbors (counterparties in interbank networks) and whether a certain threshold (such as equity) is crossed.

Our interest lies in understanding the conditions under which the initial default of a bank can trigger a global cascade of defaults, meaning that a significant fraction of banks in the network default.

\begin{figure}[h]     \centering
    \includegraphics[width=0.5\textwidth]{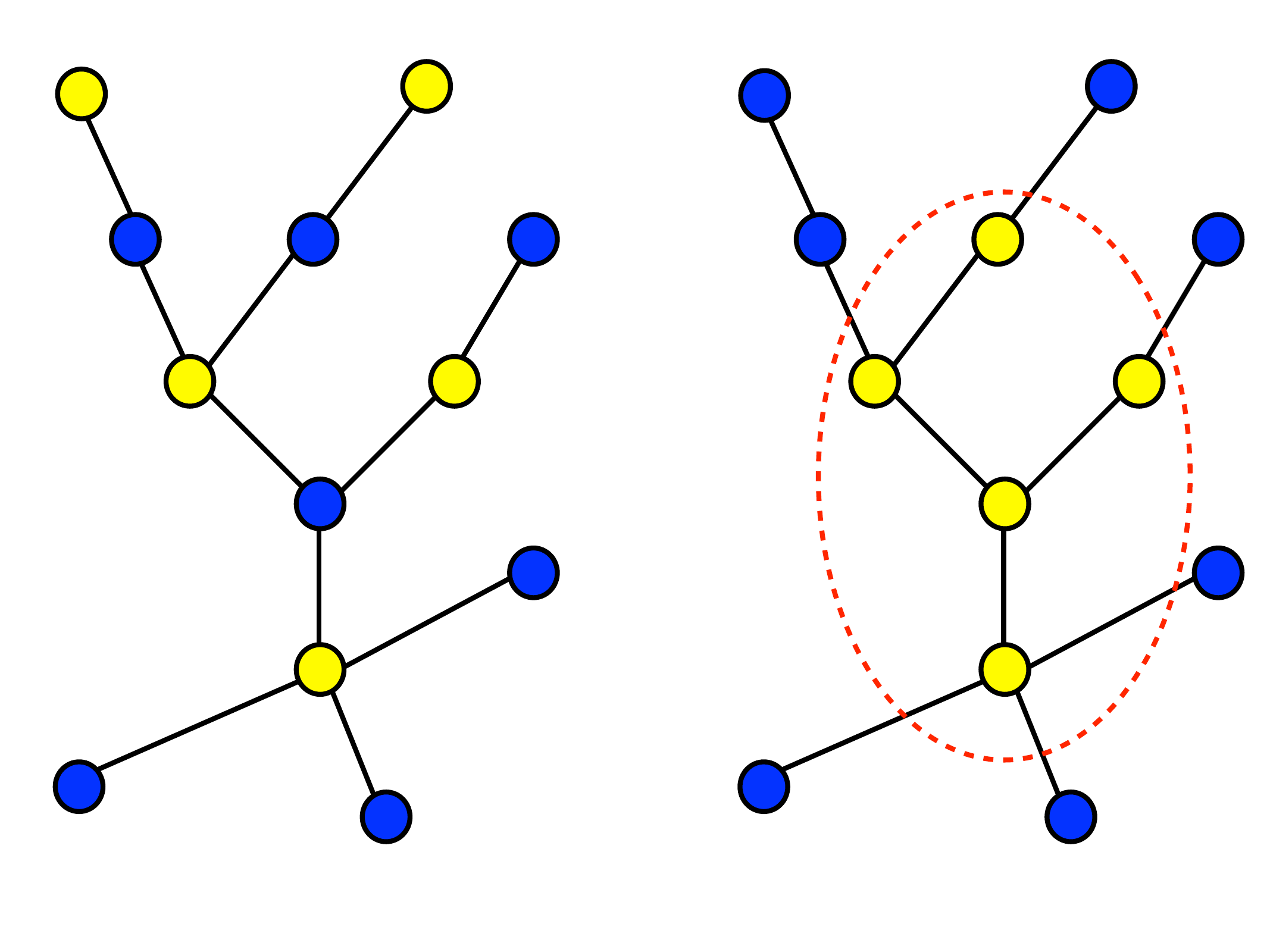}
        \caption{\footnotesize Pictorial representation of the network. Yellow nodes represent vulnerable banks. Blue nodes represent non-vulnerable ones. In the left panel, vulnerable banks are not connected between them, which prevents the occurrence of global cascades. In the right panel, global cascades can occur because vulnerable banks form a connected cluster in the network --- denoted in the picture by the red dashed line. If one vulnerable bank defaults, the entire cluster will go down.}
    \label{fig:cartoonPercolatingVulnerableCluster} 
\end{figure}

A characteristic of sufficiently sparse random networks is that they are locally tree-like, meaning there are no short loops in the network. Since the process starts with only one bankrupted bank, a chain reaction requires the network to contain banks that can default because of the failure of one counterparty only---these are referred to as vulnerable banks. Furthermore, these vulnerable banks need to be connected within the network for the cascade to spread, i.e., global cascades of bankruptcies are possible when a percolating cluster of vulnerable banks exists in the network (see  figure \ref{fig:cartoonPercolatingVulnerableCluster}).

\begin{figure}[h]     \centering
    \includegraphics[width=0.5\textwidth]{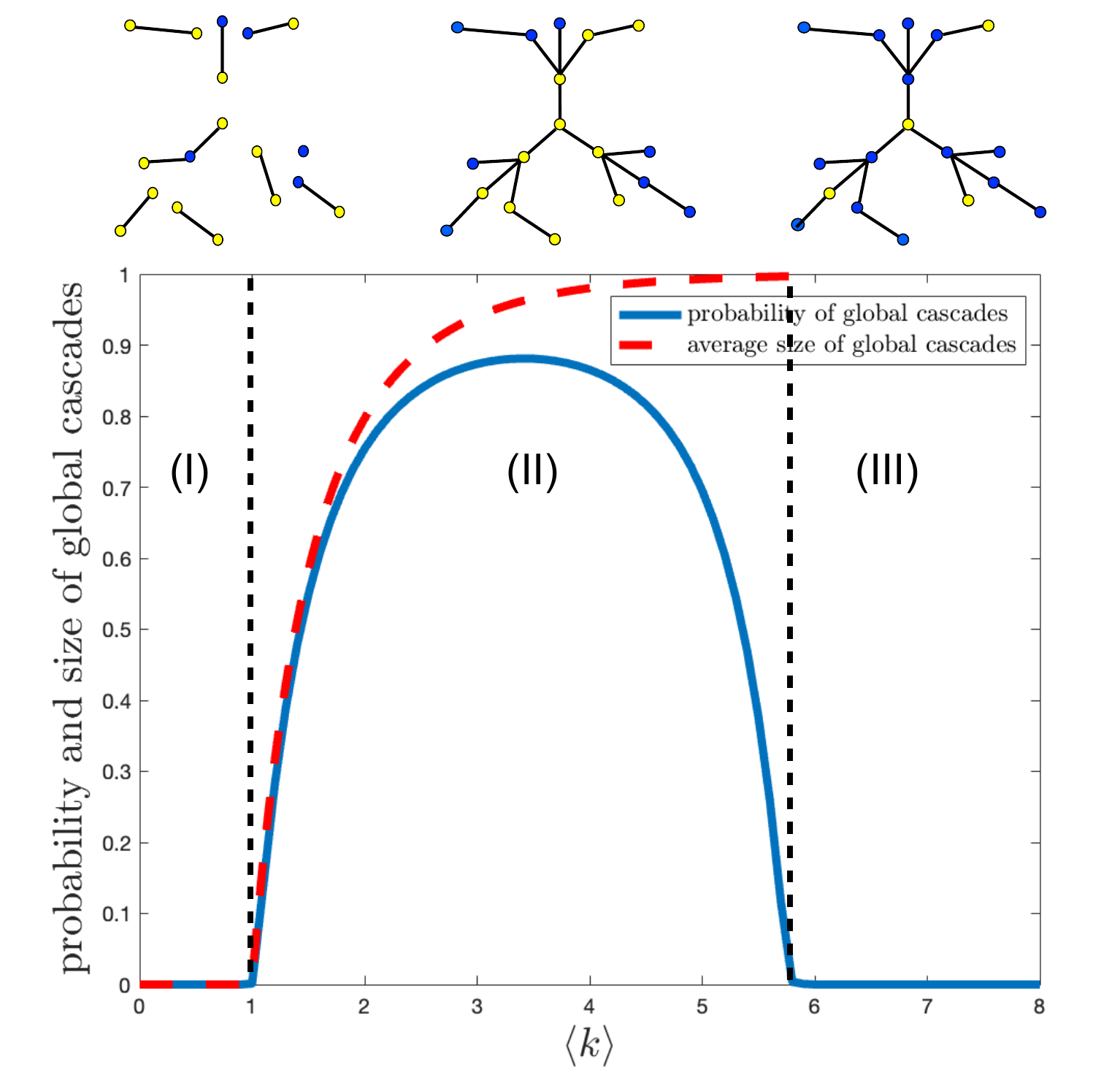}
        \caption{\footnotesize Probability of observing a global cascade on an \ER network as a function of its average degree. The activation threshold is the same for all nodes. Vertical lines mark the two transitions that separate regions where no global cascades occur from those where cascades occur with non-zero probability. In region (I), the network contains many vulnerable nodes, but it is not well-connected, so the initial activation can only propagate locally, preventing global cascades. In region (II), a percolating cluster of vulnerable nodes emerges, allowing cascades to spread through the network. In region (III), vulnerable nodes are sparse and do not form a percolating cluster, preventing cascades. The network structure is pictorially represented at the top, with three diagrams corresponding to each region. Yellow nodes indicate vulnerable nodes.}
    \label{fig:globalCascadeProbability} 
\end{figure}

Figure \ref{fig:globalCascadeProbability} shows the probability of observing global cascades as a function of the average degree.
Whether these cascades actually occur depends on whether vulnerable banks are connected to each other through paths composed exclusively of other vulnerable banks. When such a percolating cluster of vulnerable banks exists, the initial failure of one bank within this cluster can trigger the default of a large number of banks, thereby resulting in high systemic risk (see right panel of Figure \ref{fig:cartoonPercolatingVulnerableCluster} for a pictorial representation).

If, instead, vulnerable banks are scattered throughout the network and separated by non-vulnerable banks, the non-vulnerable banks effectively shield vulnerable ones from one another (see left panel of Figure \ref{fig:cartoonPercolatingVulnerableCluster} for a pictorial representation). In this configuration, cascades are unlikely to occur, and systemic risk remains low.

When the average degree is below one, the network consists of small, disconnected components without a giant component. In this case, when a bank goes bankrupt, its effect is local because it can only reach a limited number of other banks.
In the Gai-Kapadia model, the size of each interbank loan is inversely proportional to the number of loans a bank issues (the in-degree of the node).
As the average degree increases above one, a giant component emerges in the network, making banks more connected. At the same time, the number of vulnerable banks decreases as their exposures become more spread out across multiple connections. The combination of these two effects initially increases systemic vulnerability, because the presence of both a giant component and a sufficient number of vulnerable banks means that vulnerable banks can be reached from one another. As a result, the failure of one bank can propagate through the network and lead to a cascade of defaults.

However, as the average degree continues to increase, the size of each interbank exposure decreases further, leading to a reduction in the number of vulnerable banks overall. Once the average degree exceeds a critical value (the value that separates regions (II) and (III) in the example of Figure \ref{fig:globalCascadeProbability}), vulnerable banks become too dispersed within the network, making it difficult for a percolating cluster to emerge. As a result, the probability of global cascades drops, and systemic risk is significantly reduced.

From a risk management perspective, increasing diversification is beneficial because it reduces exposure to individual counterparties. Traditional risk management tools, however, tend to treat investors as isolated entities. When accounting for the interconnectedness of banks, the situation becomes more complex. Greater diversification also implies more potential pathways for shocks to propagate through the network. It is the combination of these two effects that gives rise to the non-monotonic behavior illustrated in Figure \ref{fig:globalCascadeProbability}.

In addition to the probability of observing global cascades of defaults, Gai and Kapadia study the behavior of the average size of global cascades, which appears to be monotonically increasing with the average degree. In particular, Gai and Kapadia identify a region---this is the region close to the second transition in Figure \ref{fig:globalCascadeProbability}---where the system is characterized by a robust yet fragile behavior: On the one hand, the system is robust because the probability of observing a cascade is small. On the other hand, if a cascade is triggered, the whole system is affected. From a statistical perspective, there is nothing that would distinguish between a stable and unstable system in this regime, and small fluctuations in the structure of the system can determine whether a shock is absorbed or propagated throughout the entire network.
This behavior is analogous to that discussed by Watts \cite{watts2002Simple}, and it is a good example of what Haldane referred to as the knife-edge nature of the financial system \cite{haldane2009}.

Gai and Kapadia's paper considers a relatively homogeneous system, where banks are of the same size and the network follows a Poisson degree distribution. Additionally, the network is neutral in terms of assortativity. However, as discussed earlier, real interbank networks typically do not exhibit these characteristics. Instead, they are characterized by heavy tails in the degree distribution, negative assortativity, and high clustering.

In \cite{caccioli2012heterogeneity}, the Gai-Kapadia model is studied on scale-free networks as well as Poisson networks with degree-degree correlations. Through numerical simulations, it is shown that heavy tails in the degree distribution make the system more stable against the failure of a randomly selected bank, but more fragile if the failure involves the most connected institutions. Negative assortativity, on the other hand, is found to enhance systemic stability, a result consistent with the findings in \cite{payne2009information} for the linear threshold model on undirected networks.

Rigorous analytical results for the existence of global cascades in large networks with arbitrary in- and out-degree sequences are provided in \cite{amini2012stress,amini2016resilience}. Additionally, an analytical derivation of the conditions for global cascades in assortative interbank networks is presented in \cite{hurd2017framework}.

Finally, the effect of clustering is studied in \cite{ikeda2010cascade} for the linear threshold model, where it is shown that increasing clustering makes global cascades more likely to occur.

As mentioned at the start of this section, the dynamic underlying the Gai-Kapadia model is the Furfine contagion algorithm. Beyond its theoretical exploration of financial contagion, it has also been applied to study the stability of real interbank systems \cite{furfine2003interbank,wells2004financial,degryse2004interbank,upper2004estimating,amundsen2005contagion,lubloy2005domino,van2006interbank,mistrulli2011assessing}. A common finding in the literature is that, typically, not much contagion occurs once the network is calibrated to a real-world scenario. This raises the question: Are interbank networks truly relevant? If financial networks are not as critical as initially thought, why do regulators and practitioners react with concern whenever a financial institution is at risk of going under?

In the next sections, I will explore two potential solutions to this apparent paradox. The first relates to the idea that stress can propagate within the system before actual defaults occur. The second focuses on the coupling of the interbank network with another contagion channel: contagion due to overlapping portfolios.

\section{Contagion due to credit quality deterioration}

In the threshold models considered above, shocks propagate from the borrower to the lender only after the default of the borrower. However, one could argue that the value of an interbank asset should be reduced even before the default of the counterparty. This is because the more a bank is under stress, the more it is likely that it will default, and the less its creditors expect to receive when the loan comes to maturity.

Proceeding from this intuition, and with the aim of providing a tool to estimate the build-up of systemic risk prior to the occurrence of defaults, Battiston and collaborators introduced a new algorithm, which they dubbed DebtRank\cite{battiston2012debtrank}, and which was inspired by network centrality measures such as Google PageRank.

The main assumption of the algorithm is that losses propagate from borrowers to lenders linearly: if a borrower loses x\% of its equity after a loan is issued, the lender writes down the value of the corresponding interbank asset by  x\%.
 The original algorithm further assumes that losses are propagated only once by a borrower to its creditors, meaning no further devaluation of a given interbank asset occurs after the initial write-down. This second assumption was relaxed in \cite{bardoscia2015debtrank}, whose formulation we will use for ease of exposition.

Consider a given interbank network at time $t = 0$, with $W_{ij}(0)$ denoting the amount lent by bank $i$ to bank $j$ and $E_i(0)$ its equity. If the interbank network is hit by a shock at time $t = 1$, we can write a recursive map for the relative loss $h_i(t)$ experienced by bank $i$ at time $t$: $h_i(t)=\left(E_i(0)-E_i(t)\right)/E_i(0)$. By iterating the balance sheet identity over discrete time steps and considering the linear propagation of shocks:

\be\label{eq:debtRank}
h_i(t) = {\rm min}\left\{1,\sum_j \Lambda_{ij} h_j(t) + h_i(1)\right\},
\ee   
where $\Lambda_{ij} = W_{ij}(0) / E_i(0)$, and the upper bound of $1$ ensures that lenders do not lose more than the amount they were owed. 

Equation \eqref{eq:debtRank} expresses the losses of bank $i$ as the sum of two contributions. One, $h_i(1)$, corresponding to exogenous losses, and one, $\sum_j \Lambda_{ij} h_j(t)$, coming from the network. A given counterparty $j$ contributes to the loss of $i$  
as soon as it displays an equity loss, i.e. $h_j>0$, even if the bank remains solvent (insolvency occurs when $h_j=1$).   
Furthermore, in \eqref{eq:debtRank} the contribution to $i$'s loss coming from $j$ is weighted by $\Lambda_{ij}$, which is called matrix of interbank leverage \cite{battiston2016debtrank}.

Leverage refers to the practice of borrowing money to invest, and it is typically measured as the ratio of assets to equity. Leverage is related to risk because it amplifies losses: if an investor has a leverage ratio of $\lambda$, a 1\% devaluation of its assets translates into a $\lambda\%$ equity loss. The matrix $\Lambda$, composed of elements $\Lambda_{ij}$, captures the contribution of interbank exposures to the leverage of banks within the network.

By linearizing equation \eqref{eq:debtRank}, we can compute an iterative map for the quantity $\Delta h_i(t)=h_i(t)-h_i(t-1)$, which represents the marginal losses experienced by $i$ between iterations $t-1$ and $t$ as

\be\label{eq:debtRankLinearized}
\Delta h_i(t) = \sum_j \Lambda_{ij} \Delta h_j(t).
\ee
From this, we see that $\Delta h_j = 0$ for all $j$ is a fixed point of the dynamic, whose stability depends on the largest eigenvalue $\lambda_{\text{max}}$ of the matrix $\Lambda$. If $\lambda_{\text{max}} > 1$, the fixed point is unstable, meaning that even the tiniest perturbation would lead to the default of a bank---after which the map \eqref{eq:debtRankLinearized} would no longer hold, as the upper bound in \eqref{eq:debtRank} cannot be neglected. Taking some latitude, we could say that $\lambda_{\text{max}}$ generalizes the notion of leverage to the network context: the larger $\lambda_{\text{max}}$ is, the greater the endogenous amplification of losses becomes.

Bardoscia et al. \cite{bardoscia2017pathways} study the stability of interbank networks under the DebtRank framework as these networks become more interconnected. Starting from a sparse directed acyclic graph, they randomly add links to the network while keeping the total interbank assets and liabilities of each bank constant. They observe that, on average, the system tends to become more unstable as the network's connectivity increases, a phenomenon they attribute to the emergence of particular cyclic structures within the network. The result is reminiscent of Lord May's findings that increasing the complexity of ecosystems leads to their stability \cite{may1972will}, and it provides a clear example of how an action that, according to standard models, should reduce individual risk---by increasing diversification---can, in fact, inadvertently lead to greater systemic risk. 

A generalization of DebtRank that accounts for non-linear propagation of shocks and interpolates between DebtRank and the Furfine algorithm was introduced in \cite{bardoscia2016distress}. The behavior of this non-linear model is qualitatively similar to that of DebtRank, with the system characterized by a transition between a regime where small shocks can be amplified and a regime where shocks do not propagate. However, accounting for non-linear propagation introduces a soft threshold to the propagation of shocks, offering a more realistic representation of how financial distress spreads through the system. This approach allows shocks not to be amplified for lenders when borrowers experience small losses, while still retaining the idea that losses can be transmitted before defaults occur.

Additionally, reference \cite{barucca2020network} present a framework for valuing interbank assets that incorporates both the interconnectedness of financial institutions and uncertainty in external asset values. This framework extends the DebtRank approach by focusing on how shocks and credit quality deterioration propagate through the system via mark-to-market revaluations of interbank assets. 

DebtRank has been extensively applied to assess the stability and systemic risk of financial systems across various regions using real-world empirical data \cite{silva2017vulnerability, caceres2020systemic, cuba2021network, landaberry2021contribution, leon2019short, poledna2015multi}, and it has also been used as a tool to study policy measures aimed at containing systemic risk. In particular, Poledna and Thurner \cite{thurner2013debtrank, poledna2016elimination} consider an agent-based model of the interbank system coupled with the economic system. In the model, firms approach banks for capital to fund investments, and banks establish interbank connections to ensure the flow of such funds. Depending on their production outcomes, some firms may go bankrupt, passing their losses to the financial system.

Poledna and Thurner measure the frequency and size of cascades in the interbank market as the model runs, and they propose introducing a tax based on the contribution to systemic risk that a new interbank loan imposes on the system, measured using DebtRank. By comparing the benchmark of no tax with the scenario where the tax is implemented, they show that it is possible to reduce the probability of large cascades occurring in the interbank market while maintaining the same level of credit for the economy. In contrast, a tax based purely on the size of the interbank loan (without considering its contribution to systemic risk) leads to a reduction in available credit.

\section{Contagion due to overlapping portfolios}

I have so far discussed the case of contagion through direct linkages between financial institutions. As mentioned earlier, contagion can also occur  when financial institutions invest in common assets. In this case, while there is no direct contract between the institutions, shocks can still propagate from one to another indirectly through asset prices. If one institution is under stress and needs to sell part of its assets, those assets may be devalued due to market impact---the tendency of prices to respond to trading activity \cite{bouchaud2009markets}. This devaluation leads to mark-to-market losses for other institutions holding the same assets. In turn, these institutions may be forced to liquidate their own assets in response to their losses, further depressing prices and potentially triggering a fire sale.

\begin{figure}[h]     \centering
    \includegraphics[width=0.8\textwidth]{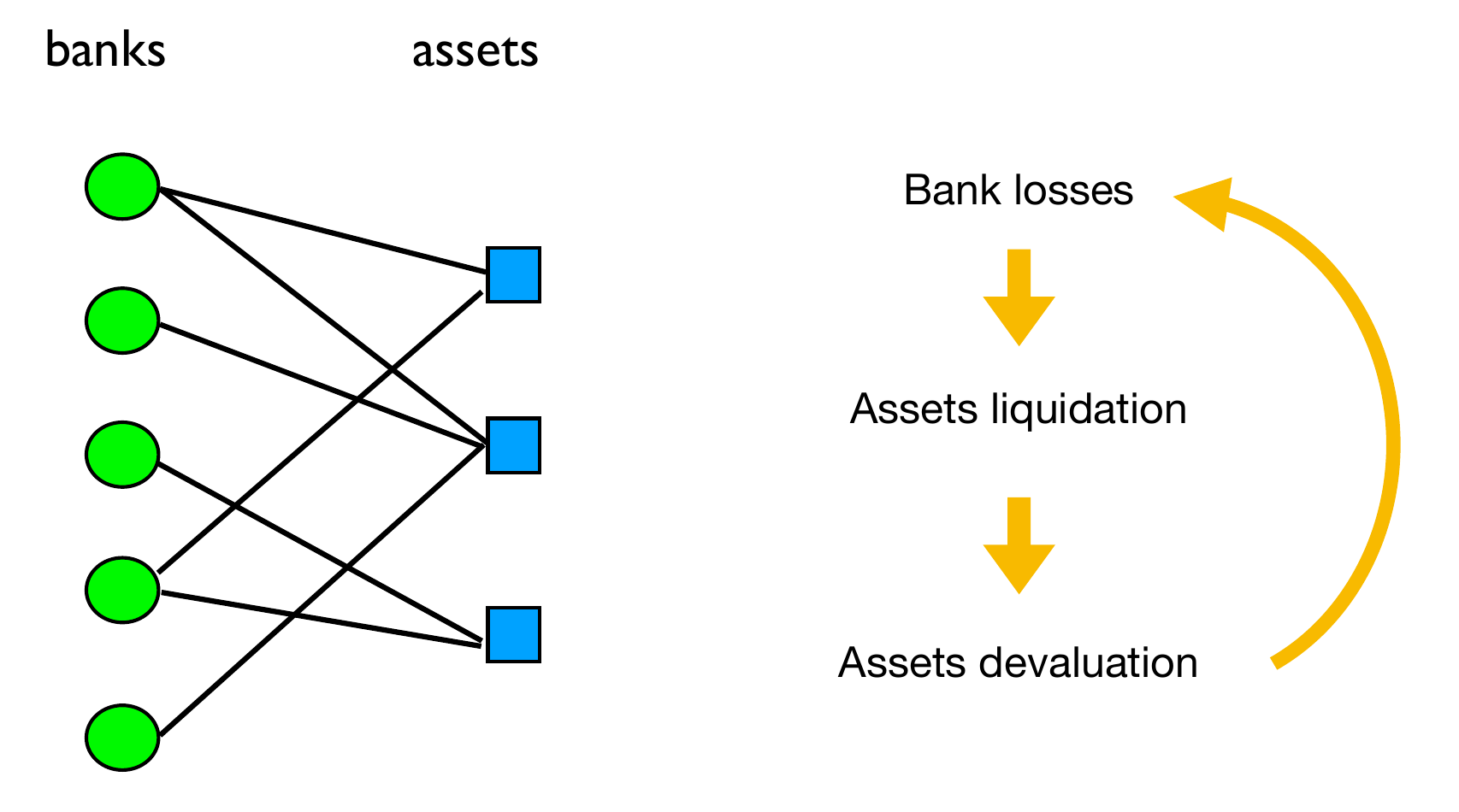}
        \caption{\footnotesize {\bf Left panel:} Stylized representation of a network of overlapping portfolios. Banks interact indirectly through their common assets. {\bf Right panel:} feedback mechanism of contagion due to overlapping portfolios.}
    \label{fig:cartoonBipartite} 
\end{figure}

Network models can be adapted to represent this contagion mechanism (see figure \ref{fig:cartoonBipartite}). Instead of focusing on direct interbank connections, the network becomes a bipartite network of banks and assets that represents banks' portfolios. In this structure, banks interact indirectly through their overlapping portfolios, i.e., the common assets they hold.

Network models of contagion due to overlapping portfolios were  studied in \cite{caccioli2014stability}, where the authors introduced two approaches: a linear threshold model, in which banks remain passive until they become insolvent and then liquidate their entire portfolios, and a model where banks actively manage risk, responding to losses by selling off part of their portfolios to reduce exposure. In both cases, asset prices respond to liquidation through a linear market impact on log-prices

\be
\log \left(p_a(t)/p_a(0)\right)= \eta_a q_a(t),
\ee
where $p_a(t)$ is the price at time $t$, and $q_a(t)$ is the volume of asset $a$ that has been liquidated up to time $t$, and $\eta_a$ is a market impact parameter that is associated with the asset's liquidity, that is its sensitivity to trading.

Similarly to the analysis of Gai and Kapadia, the key question is understanding the conditions under which a global cascade of defaults can be triggered by the initial default of a bank or the devaluation of an asset. As in the previous section, the threshold model can be understood as a network percolation problem, where a vulnerable bank is one that can default due to the devaluation of a single asset. An equivalent description can be provided using a branching process, which is the one given in \cite{caccioli2014stability}. In a branching process, a progenitor generates a number of offspring $n$ drawn from a distribution $P(n)$, and each offspring in turn produces more offspring according to the same distribution $P(n)$. The question is whether the process will eventually die out or if there is a non-zero probability it will continue indefinitely. If the expected number of offspring per individual is $\overline{n} = \sum_n n P(n)$ and $\overline{n} < 1$, the population will go extinct with probability one. If, however, $\overline{n} > 1$, there is a chance the population will survive. This is analogous to epidemiology, where $\overline{n}$ is equivalent to the basic reproduction number.

The analogy with financial contagion is clear. Given a generation 0 defaulted bank, it can trigger a certain number of generation 1 defaults among banks that share common investments, which can in turn trigger generation 2 defaults. If the expected number of next-generation defaults per current bankrupted bank is larger than one, the contagion can continue to spread through the network with a non-zero probability. In reality, the situation is more complex because banks differ in size, investments, leverage, and other factors. However, the result can be generalized by introducing a matrix $B$, where each element $B_{ij}$ represents the probability that the default of bank $j$ will trigger the default of bank $i$ due to their common investments. The matrix $B$ can be estimated from data about portfolio holding and the market impact parameters, and its largest eigenvalue plays the role of the basic reproduction number in this case, further strengthening the analogy between financial contagion and epidemiology.

In \cite{caccioli2014stability}, the behavior of the model is studied on the bipartite equivalent of \ER random graphs using the same balance sheet setting as Gai and Kapadia. Similar behaviors to those observed for counterparty default risk are found, particularly regarding both the probability of global cascades and the average size of those cascades. The case of heavy-tailed degree distributions is studied via numerical simulations in \cite{banwo2016effect},  while an analytical solution of a simplified model is given in \cite{onaga2023financial} using techniques developed in \cite{fennell2019multistate} for multistate dynamical processes on networks.

A bipartite network threshold model for contagion due to overlapping portfolios is also introduced in \cite{huang2013cascading}, where it is calibrated using balance sheet data from US commercial banks in 2007. The paper shows that abrupt transitions in the number of defaults occur as a function of asset liquidity and the size of the external shock (devaluation of an asset class). Furthermore, they validate their model through an analysis of true positives and false positives for defaults observed in the US from 2008 to 2011, demonstrating that the network model has some explanatory power.

While the threshold model is a useful benchmark for understanding how network structure affects stability, in practice, banks actively manage their own risk by reducing investments before becoming insolvent in an attempt to contain their losses. In \cite{caccioli2014stability}, a simple protocol for such preventative liquidation, based on leverage targeting, is considered. When a bank suffers a loss, its leverage---and therefore its risk---increases. In response, the bank may liquidate part of its assets to bring its leverage back to the desired level.

However, reference \cite{caccioli2014stability} demonstrates that preemptive liquidation actually makes the system more unstable compared to the benchmark of a passive investor. Moreover, the more aggressively a bank tries to reach its leverage target, the more unstable the system becomes. The reason is intuitive: when banks liquidate assets to reduce leverage, they push prices down, leading to further losses. In the absence of activation thresholds for liquidation, this makes it easier for a downward spiral to take hold. In fact, as shown in \cite{caccioli2012impact}, it is even possible for an individual investor using leverage to drive themselves into bankruptcy by attempting to exit their position. The presence of a network further exacerbates this tendency.

Stylized models of contagion, such as those discussed above, are useful for understanding the mechanics of the system and how its response to shocks is influenced by structural parameters (such as connectivity) and dynamical ones (such as the liquidity of assets). Beyond theoretical analyses, more detailed frameworks have been developed and calibrated to real systems. A leverage targeting fire-sale model was introduced and applied to European banks in \cite{greenwood2015vulnerable}, and further extended in \cite{duarte2021fire}, where a study of the US banking system is presented. Cont and Schaanning \cite{cont2017fire} introduced a hybrid model, where banks remain passive below a certain threshold but switch to leverage targeting once the threshold is crossed. They calibrated this model to bond holdings of EU banks. Notably, they also introduced the concept of indirect exposure, accounting for additional risk arising from the network of overlapping portfolios. In related work, they introduced an index to monitor indirect contagion, derived from the spectrum of the matrix of liquidity-weighted overlapping portfolios \cite{cont2019monitoring}. 

Historically, research on systemic risk focused primarily on banks, as they were regulated by central banks, where part of the early literature emerged, and because, albeit sparse, data were available for them. When considered collectively, empirical studies on banking systems suggest that the contagion effect of overlapping portfolios is larger than that of interbank lending. More recently, research has expanded to model other types of financial institutions. For instance, \cite{barucca2021common} provides an empirical analysis of the network of overlapping portfolios between European investment funds and UK-regulated banks and insurance companies, while \cite{fricke2021vulnerable,fricke2023connected} adapt the model of \cite{greenwood2015vulnerable} to study the case of funds.

While from a technical perspective, the dynamics in models of funds, such as leverage targeting, may resemble those observed for banks, the underlying mechanisms behind these rules differ. It is crucial that these distinctions are modeled as realistically as possible to ensure accurate calibration to real data.

Additionally, new system-wide stress-testing frameworks have been proposed in the literature to model various institutions, such as banks, funds, and insurers \cite{farmer2020foundations, caccioli2024modelling}. These studies demonstrate that neglecting intersectoral overlapping portfolios can result in a severe underestimation of systemic risk.

\section{Multiple contagion channels}

In the previous sections, we considered two channels of financial contagion: contagion due to counterparty default risk and contagion due to overlapping portfolios. However, as discussed earlier, banks simultaneously interact through different kinds of relationships. The multilayer nature of financial networks has been explored in \cite{poledna2015multi,bargigli2015multiplex}. In practice, multiple contagion channels are often active simultaneously, making it essential to study their combined effects. This has been explored to some extent in works such as \cite{cifuentes2005liquidity, gai2010contagion, nier2007network}, which examine scenarios where all banks interact in an interbank network and invest in the same asset. The asset dynamics follow a similar pattern to what was discussed in the previous section. Their findings indicate that including these additional interactions clearly leads to additional losses compared to the benchmark scenario of only counterparty default risk. May and Arinamympathy \cite{may2010systemic} take this further by considering interactions between asset classes in a stylized setting.

Caccioli et al. \cite{caccioli2015overlapping} attempt to address the question raised at the end Section \ref{sec:counterpartyRiskContagion}, namely, why regulators are concerned about bank defaults if contagion in interbank lending networks is not typically significant. The authors study the Austrian interbank network, assuming a certain degree of correlation between banks' external assets. They perform stress tests using the Furfine algorithm, accounting for the devaluation of external assets upon liquidation under three different scenarios: (i) contagion occurs only due to counterparty default risk, (ii) contagion occurs only due to overlapping portfolios, and (iii) contagion occurs due to both channels simultaneously.

They find that, if the correlation of external assets is large enough, while each contagion channel may not result in large cascades when considered individually, the combination of the two can lead to systemic instability. This suggests that while the interbank lending network may not be a significant risk factor on its own, it can play a crucial role in amplifying contagion arising from overlapping portfolios.

Using data on interbank exposures and portfolio holdings of Mexican banks, Poledna et al. \cite{poledna2021quantification} generalize DebtRank to multilayer networks to study their combined effect, and they find that neglecting the interaction between the two contagion channels significantly underestimates systemic losses.
Building on the idea of interacting contagion channels, Wiersema et al. \cite{wiersema2023scenario} develop an eigenvalue-based approach, demonstrating that the instability caused by interacting channels can far exceed the sum of the instabilities from individual channels acting in isolation.

All in all, this line of research makes it clear that to properly estimate and manage systemic risk, one must account for as many contagion channels as possible simultaneously. Neglecting their interaction can lead to significant underestimation of systemic losses, as the combined effect of these channels can be much greater than the sum of their individual impacts. 

\section{Dynamical models}

The models discussed so far are essentially static, where a network is given, a shock is applied, and the system evolves according to a deterministic dynamic. These models are highly useful for understanding how the structural constraints of the network and simple behavioral rules of financial institutions can lead to the endogenous amplification of shocks. Moreover, they have practical applications in stress testing, where the focus is on analyzing scenarios that unfold over a relatively short time period, during which the network structure can be considered effectively fixed.

However, to fully understand how a system can endogenously transition from a normal state to a crisis regime, we need dynamic models in which investments and connections evolve over time. This is where agent-based modeling becomes particularly valuable \cite{axtell2022agent}. In agent-based models, individual financial institutions (agents) interact with one another based on a set of behavioral rules, allowing for the emergence of complex system-wide phenomena. By simulating the evolution of investments, network structures, and decision-making processes, agent-based models can offer a more realistic representation of how systemic risk develops and propagates over time.
I report here a few results obtained using simple agent-based models, which clearly demonstrate how standard financial risk management tools can, under certain conditions, lead to systemic instabilities.

A dynamical model of overlapping portfolios that includes stochasticity is introduced and studied in \cite{corsi2016micro}. In this model, banks actively manage their risk through leverage targeting, and asset prices fluctuate over time due to both the trading activity of banks and random fluctuations. The model shows that financial innovation, through a reduction in the cost of diversification, can trigger a transition from a stable market regime to one characterized by bubbles and bursts.

Thurner et al. \cite{thurner2012leverage} consider a dynamical model of leveraged investment funds. In this model, funds borrow money from banks and use leverage to purchase undervalued assets, providing a stabilizing mechanism for the market, which is also populated by noise traders. However, the bank requires them to keep their leverage below a specified cap, and they are subject to margin calls when this limit is breached. A margin call forces the fund to repay part of the borrowed money. If a fund is fully leveraged, even a small market fluctuation can trigger a margin call, forcing the fund to sell some of its assets to meet its creditors' demands. This, in turn, depresses the asset price further, potentially triggering more margin calls and creating a downward spiral. Thurner and coauthors demonstrate that this mechanism can explain the emergence of heavy tails and clustered volatility in financial markets.

Using the model from \cite{thurner2012leverage} as a benchmark, Poledna et al. \cite{poledna2014leverage} study the effect of two different regulatory schemes for banks. The first is the one mandated by Basel II, where exposures are mitigated solely by collateral, and the second is a perfect-hedging scheme, where banks are additionally required to hold options to hedge their exposures to funds. Interestingly, they find that both policies increase systemic stability when leverage is low, but reduce it when leverage is high. This is because these policies lead to increased synchronization among funds that need to deleverage, inadvertently increasing systemic risk. This is another example of how measures intended to reduce risk can sometimes exacerbate systemic risk.

The impact of Basel II regulations is also central to the work of Adrian and Shin \cite{adrian2010liquidity}), who explore how a Value-at-Risk (VaR) constraint, such as that required by Basel II, leads to leverage procyclicality. This phenomenon causes banks to increase their leverage during favorable market conditions (low volatility and rising prices) and reduce it during downturns (high volatility and falling prices).

From an interacting systems perspective, this procyclicality creates a feedback loop that can fuel market bubbles and crashes. In periods of high prices and low volatility, banks increase their leverage and expand their investments, driving prices even higher and reducing volatility further. This positive feedback loop promotes continued balance sheet expansion. Conversely, during market downturns, banks sell off assets, pushing prices down and increasing volatility, which prompts even more asset liquidations, reinforcing the negative spiral.

A simple agent-based model illustrating leverage procyclicality is presented by Aymanns et al. \cite{aymanns2015dynamics}. Their model demonstrates how endogenous market oscillations---characterized by gradual increases in leverage and asset prices followed by sudden crashes---emerge when banks manage risk based on historical volatility estimates to determine their target leverage. These oscillations occur within a chaotic regime of the system. In an extended version of the model, which includes a leverage-targeting bank and an unleveraged fundamentalist investor, Aymanns et al. \cite{aymanns2016taming} show that, when roughly calibrated with realistic parameters, the model can replicate leverage cycles lasting 10 to 15 years, akin to those observed in real markets.

Aymanns et al. \cite{aymanns2016taming} also investigate the effects of different capital buffer policies, ranging from procyclical to countercyclical measures, and find that the optimal policy depends on the size of the banking system. Basel II's procyclical rules are found to be effective when the banking system is small and leverage is low, as in such a case the bank's impact on the market is limited, making the approximation of it as an isolated agent implicit in standard risk management tools reasonable. However, when the banking system is large and leverage is high, maintaining constant leverage becomes the optimal strategy, as Basel II's procyclical rules tend to destabilize the system.  Finally, the authors showed that slowing down the response of the bank, i.e. giving the leveraged investor more time to pay the loan back, is in fact the most effective strategy to improve systemic stability.
\section{Conclusions}

This chapter reviewed complex systems approaches to the study of financial systemic risk, an area that has seen significant research activity from complexity scientists since 2008.

I discussed how domino effects in financial networks can be understood through the emergence of a percolating cluster of vulnerable banks, or equivalently, as a supercritical branching process. Additionally, I examined simple agent-based models whose dynamics can endogenously lead to market instability. I also highlighted several examples where actions intended to reduce individual risk, according to standard risk management tools, can inadvertently increase systemic risk.

Network models of financial contagion began as simple, stylized models designed to gain intuition about how shocks spread across banks and how network structure might influence this propagation. Over time, these models have evolved to account for more complex contagion mechanisms, incorporate institutions beyond banks, and introduce more realistic dynamics. This evolution has paved the way for the development of network-based macroprudential stress-testing frameworks, which can be calibrated with real-world data and complement traditional microprudential stress testing, where the interconnections between financial institutions and feedback loops are often overlooked.

Clearly, the literature on systemic risk is much broader than what we could cover here. Key topics that have emerged include network reconstruction \cite{squartini2018reconstruction, anand2018missing}, liquidity contagion \cite{anand2012rollover, soramaki2013sinkrank,cimini2016entangling}, the estimation of systemic risk and connectivity from market data \cite{brownlees2017srisk, billio2012econometric, aste2021stress}, and the development of stress-testing frameworks that incorporate climate risks and estimate transition risks \cite{battiston2017climate,monasterolo2020climate}. For a more comprehensive understanding of financial networks and contagion, I also recommend the following reviews: \cite{glasserman2016contagion, bardoscia2021physics}.

In the broader field of complexity applied to finance, it is also worth mentioning the literature on minority games \cite{challet1997emergence,challet2004minority} in Physics, as well as the literature on heterogeneous agents in Economics \cite{brock1998heterogeneous}, which have contributed to understanding potential sources of market instability (see for example \cite{brock2009more,marsili2014complexity,bardoscia2012financial,caccioli2009eroding}).

Beyond its scientific value, the primary motivation for managing systemic risk is to mitigate its negative consequences on the real economy. The financial system's role should be to ensure that funding is properly allocated to firms and households. When a financial crisis occurs, it can spill over to the real economy, causing rising unemployment, lower production, and GDP loss. Therefore, models of financial systemic risk must eventually be coupled with macroeconomic models, such as those recently developed in \cite{hommes2022canvas,poledna2023economic}, where agent-based modeling approaches have already proven fruitful \cite{gatti2011macroeconomics,tedeschi2012bankruptcy,gualdi2015tipping,gualdi2015endogenous,geanakoplos2012getting,baptista2016macroprudential}. However, these models still lack a direct connection to financial models, making their integration an important area for future research.

In conclusion, complexity science has significantly advanced our understanding of systemic risk by providing tools to model the intricate interconnections and feedback loops that shape financial systems. As the field progresses, future research must continue to integrate these approaches with macroeconomic modeling and policy frameworks to create actionable insights for regulators and policymakers. By doing so, we can better prepare for the risks posed by financial crises and help safeguard the stability of the real economy.

\section*{Acknowledgments} I would like to thank J. Doyne Farmer, R. Maria del Rio-Chanona, and Marco Pangallo for inviting me to write this chapter, and J. Doyne Farmer for his valuable comments.

\bibliographystyle{unsrt}
\bibliography{chapterSystemicRisk}

\end{document}